\documentstyle[preprint,seceq,eclepsf]{jpsj}
\def\NY{\nonumber \\}
\def\dfrac#1#2{{\displaystyle\frac{#1}{#2}}}

\def\d{{\rm d}}
\def\i{{\rm i}}
\def\e{{\rm e}}

\def\E{\mbox{\boldmath $E$}}
\def\r{\mbox{\boldmath $r$}}
\def\k{\mbox{\boldmath $k$}}
\def\A{\mbox{\boldmath $A$}}
\def\j{\mbox{\boldmath $j$}}
\def\x{\mbox{\boldmath $x$}}
\def\y{\mbox{\boldmath $y$}}
\def\mPi{\mbox{\boldmath  $\mit \Pi$}}
\def\iPi{{\mit \Pi}}

\def\ee{e^\ast}

\def\re{{\rm Re}}
\def\im{{\rm Im}}
\def\Dt{\dfrac{\partial}{\partial t}}
\def\Dkx{\dfrac{\partial}{\partial k_x}}

\def\ko{(\k,\omega)}
\def\kop{(\k^\prime,\omega^\prime)}

\def\runtitle{
Dynamical Induction of $s$-wave Component in $d$-wave Superconductor
}
\def\runauthor{Atsuya {\sc Kumagai} and Hiromichi {\sc Ebisawa}}

\title
{
Dynamical Induction of $s$-wave Component in $d$-wave Superconductor \\
Driven by Thermal Fluctuations \\
} 

\author
{ 
Atsuya {\sc Kumagai} \footnote{E-mail: kumagai@cmt.is.tohoku.ac.jp}
and Hiromichi {\sc Ebisawa} \footnote{E-mail: ebi@cmt.is.tohoku.ac.jp}
}

\inst
{
Graduate School of Information Sciences, Tohoku University, Sendai 980-8579
}

\recdate
{
\today
}

\abst
{
We investigated the mutual induction effects between the $d$-wave 
and the $s$-wave components of
order parameters due to superconducting 
fluctuation above the critical temperatures and calculated its
contributions to paraconductivity and excess Hall conductivity based on the
two-component stochastic TDGL equation.
It is shown that the coupling of two components increases paraconductivity
while it decreases excess Hall conductivity compared to the cases when each 
component fluctuates independently.
We also found the singular behavior in the paraconductivity
and the excess Hall conductivity dependence on the coupling parameter
which is consistent with
the natural restriction among the coefficients of gradient terms.
}

\kword
{
superconducting fluctuation, TDGL theory, two-component superconductor,
paraconductivity, excess Hall conductivity
}

\begin{document}
\sloppy
\maketitle

\section{Introduction}
Recently many experimental results indicate the mixed pairing symmetry
of Cooper pairs in high-$T_{\rm c}$ cuprates which consists of
$d$- and $s$-wave-like components~\cite{sri,will}. 
Theoretically, it is shown
that certain classes of microscopic models
lead to dominant $d$-wave pairing with 
subdominant $s$-wave pairing~\cite{Feder}.
Macroscopic properties of superconductors of these
mixed symmetry can be studied by means of the Ginzburg-Landau (GL)
thoery for two-component order parameter.
Phenomenologically, Joynt~\cite{Joynt} first introduced the 
GL free energy of two-component superconductors
based on the group theoretical consideration~\cite{Sigrist}.
Microscopic derivations of the corresponding two-component GL equations 
have been performed by several authors~\cite{Xu,Xu2,Feder,Han}.

Due to those macroscopic arguments, it is shown that,
in tetragonal lattice,
the coexistence of the two components is not expected in the bulk,
but the subdominant order parameter is induced
by spatial variations
because two components are coupled through ``mixed gradient'' term;
this effect has been intensively investigated under the
circumstances such as vortices~\cite{Franz,Han,Alv,Chang}, 
impurities~\cite{Xu2}, currents~\cite{Zap} etc., thereby their anisotropic
features being clarified.
Such inductions have been also studied from more microscopic 
standpoints~\cite{kuboki,ichi,hime}.
Nonetheless there is so far no considerations of the effect beyond these
static cases.
There should be {\it dynamical} situations where the
order parameter varies spatially so that
the subdominant component is induced.
We now point out this possibility, which is driven
by the superconducting fluctuaion above critical temperature,
and give an estimation of the effect to the paraconductivity
and the excess Hall conductivity.
It should be stressed that, within our considerations, 
the origin of the induction is superconducting fluctuation itself,
unlike the recent studies treating vortex states based on
two-component TDGL equation~\cite{Alv,Chang}, where the induction is 
attributed to intrinsic structures of vortices.

The conduction process which contributes to the enhancement of 
electrical conductivity due to superconducting fluctuaion is first 
pointed out by Aslamasov and Larkin(AL)~\cite{AL}.
Besides, there is a prediction about the behavior of Hall conductivity 
resulting from AL process~\cite{FET}, though it is not yet clear
whether it accounts for the Hall anomaly~\cite{naga}.
AL process can also be described 
within the framework of Time-Dependent Ginzburg-Landau(TDGL) equation
and several formulations have been done~\cite{Schmid,Dorsey,dami}
based on the stochastic TDGL equation.

In this work, based on the stochastic two-component TDGL equation,
we show our calculation for the paraconductivity and
the excess Hall conductivity and discuss the importance
of the induction effect of the subdominant component
through the superconducting fluctuation.
Since we limit the considerations to the case of weak magnetic field,
we will make the simple linear or bilinear
expansion of the response current
in terms of the external electric and magnetic field.

The characteristic features of the induction effect we will show
is firstly that no anisotropy arises in spite of the anisotropy
in GL free energy and secondly
that the induction reduces the excess Hall conductivity
while it enhances the paraconductivity.
The results will show the singular behavior of these transport
coefficients in increment of the coupling of two components
through the spatial variation.
This is direct reflection of the stability condition of
the normal state against the superconduting fluctuation. 

In Sec.\ref{sec2},
we first give the formulations of fluctuation conductivity tensor
based on the two-component
TDGL equation as a stochastic differential equation.
In Sec.\ref{sec3}, we show the numerical results 
and see how the coupling of two components
affects them. In Sec.\ref{summary}, we summarize the results
and give some discussions.
In this paper we set
$\hbar = k_{\rm B} = c =1.$

\section{Formulation for the stochastic TDGL equation} \label{sec2}
In this section we show the formulation for the 
paraconductivity and excess Hall conductivity
based on the stochastic two-component TDGL equation.
In the following, electric and magnetic field is expressed 
as scalar and vector potential
$\phi(\r),\A(\r)$, respectively.
The charge of a Cooper pair is $e^* = -2e < 0$. 
We introduce a differential operator
$\mPi = -\i \nabla - e^* \A(\r).$

We start with the two-component TDGL equation
\begin{subequations}
\begin{equation}
     \gamma_d \left( \Dt + \i \ee \phi \right) d
  = -\left( \frac{\mPi^2}{2 m_d} + \alpha_d \right) d
          - \frac{(\iPi_y^2 -\iPi_x^2)}{2 m_v} s + f_1
\end{equation}
\begin{equation}
     \gamma_s \left( \Dt + \i \ee \phi \right) s
  = -\left( \frac{\mPi^2}{2 m_s} + \alpha_s \right) s
          - \frac{(\iPi_y^2 -\iPi_x^2)}{2 m_v} d + f_2
\end{equation}
\label{TDGL}
\end{subequations}
\begin{eqnarray}
  \j(\r) &=& \frac{e^*}{m_d} (d^* \mPi d + {\rm c.c.}) 
          +\frac{e^*}{m_s} (s^* \mPi s + {\rm c.c.}) \NY
    & & -\hat{\x}\frac{e^*}{m_v} (d^* \iPi_x s + s^* \iPi_x d + {\rm c.c.}) 
        +\hat{\y}\frac{e^*}{m_v} (d^* \iPi_y s + s^* \iPi_y d + {\rm c.c.}) 
  \label{current}
\end{eqnarray}
which is based on Joynt's two-component GL free energy~\cite{Joynt} and
also can be derived from such lattice models as $t$-$J$ model~\cite{spar}.
$d$ and $s$ mean superconducting order parameters of
$d$-wave and $s$-wave component, respectively. Both GL coefficients
$\alpha_d$ and $\alpha_s$ are positive because the system is above the
transition temperatures.
We added the white noise source terms $f_1,f_2$ as representations of 
thermodynamical fluctuations, which make the
mean square expectation value of order parameter finite 
even above the transition temperature.
It is the main characteristics of the TDGL equation that the
two components are coupled
via so-called ``mixed gradient'' terms. The main purpose of this
paper is to clarify the effects of this term on the electric
transport properties, and this effects are represented as the
coefficient $m_v^{-1}$. 
Because of this anisotropic term we should have set electrical field
as $\E = ( E_x , E_y ,0)$ for more generality, but it will be
found later that this anisotropy is cancelled with that of current and
no anisotropy arises in the fluctuation conductivity tensor.
For this reason we set $\phi(\r) = -Ex , \A(\r) = Bx\hat{\y}$.

First, we rewrite the TDGL equation(\ref{TDGL}) into the simpler form.
Keeping up to bilinear terms in external fields,
Fourier transformed TDGL equation becomes
\begin{equation}
  \i \gamma \omega \psi \ko
  = - \left( \eta_k + \gamma V_k \Dkx \right) \psi\ko + f\ko, \label{TDGL2}
\end{equation}
\begin{equation}
V_k = \ee (E- \i B \gamma^{-1} k_y W_y )
\end{equation}
where 
\begin{equation}
\psi = \pmatrix{ d \cr s \cr},
\end{equation}
\begin{equation}
  f = \pmatrix{ f_1 \cr f_2 \cr},
\end{equation}
\begin{equation}
\eta_k =  \frac{k_x^2}{2} W_x + \frac{k_y^2}{2} W_y + \alpha,
\end{equation}
\begin{equation}
  W_x=\pmatrix{ m_d^{-1} & -m_v^{-1} \cr
               -m_v^{-1} &  m_s^{-1} \cr},
\end{equation}
\begin{equation}
  W_y=\pmatrix{ m_d^{-1} &  m_v^{-1} \cr
                m_v^{-1} &  m_s^{-1} \cr},
\end{equation}
\begin{equation}
  \alpha = \pmatrix{ \alpha_d & 0 \cr 0 & \alpha_s \cr},
\end{equation}
\begin{equation}
  \gamma = \pmatrix{ \gamma_d & 0 \cr 0 & \gamma_s \cr}.
\end{equation}
Then we can formally express the solution of the TDGL equation:
\begin{equation}
  \psi\ko = G\ko f\ko
\end{equation}
where
\begin{equation}
G\ko \equiv
  \left( i \gamma \omega + \eta_k + \gamma V_k \Dkx \right)^{-1}.
\end{equation}

Here we assume the correlation function between white noises as
\begin{equation}
\langle f(\k',\omega') f^\dagger (\k,\omega) \rangle 
= a \delta_{\k \k'} \delta_{\omega \omega'}, \label{ff}
\end{equation}
where the coefficient $a$ is $2 \times 2$ matrix and should be 
determined from the linearized two-component GL free energy
\begin{equation}
  F_k = \psi(\k)^\dagger \eta_k \psi(\k). \label{GLF2}
\end{equation}
$\psi(\k)$ means the order parameter of wave vector $\k$ 
at the specific time.
Corresponding distribution function is
\begin{equation}
  Z_k = \int {\rm e}^{-F_k /T} \d d_1 \d d_2 \d s_1 \d s_2
      = \frac{( \pi T)^2 }{ \det \eta_k},
\end{equation}
where indices 1 and 2 mean real and imaginary part, respectively.
This distribution function gives rise to
\begin{equation}
  \langle \psi(\k) \psi(\k)^\dagger \rangle _0
   = -T \pmatrix{ \frac{\partial}{\partial \eta_d} 
                    & \frac{1}{2}\frac{\partial}{\partial \eta_v} \cr
           \frac{1}{2}\frac{\partial}{\partial \eta_v} 
                    & \frac{\partial}{\partial \eta_s} \cr}
       \ln Z_k = T \eta_k^{-1}.
\end{equation}
$\langle \cdots \rangle _0$ means thermal average in the absence of
external fields.
The stochastic TDGL equation(\ref{TDGL}) also realizes this 
equilibrium values, provided we set 
\begin{equation}
  a =  T ( \gamma + \gamma^\dagger ). \label{aa}
\end{equation}
In fact, from eq.(\ref{ff}) with eq.(\ref{aa}), we obtain
\begin{eqnarray}
  \langle \psi(\k) \psi(\k)^\dagger \rangle _0
   &=& \sum_\omega \langle \psi \ko \psi^\dagger \ko \rangle _0 \NY
   &=& \sum_\omega G_0 (\k,\omega)
    \langle f(\k,\omega) f^\dagger(\k ,\omega) \rangle 
    G_0 ^\dagger(\k,\omega) \NY
   &=& T  \int_{-\infty}^{\infty} \frac{\d \omega}{2 \pi \i}
      [(\omega - \i \gamma^{-1} \eta_k )^{-1} \eta_k^{-1}
    - \eta_k^{-1} (\omega + \i \eta_k \gamma^{-1 \dagger} )^{-1}] \NY
   &=& T \eta_k^{-1}
\end{eqnarray}
where
$G_0\ko = (i \gamma \omega + \eta_k)^{-1}$ stands for the Green's
function of TDGL equation(\ref{TDGL2}) in the absence of external fields.

Next we perform expansions with respect to external fields $V_k$.
Resolvent $G\ko$ is expanded as follows:
\begin{equation}
   G\ko = G_0\ko + G_1\ko + G_2\ko + \cdots . \label{resolve}
\end{equation}
Note that, in the above expansion, each term acts on any function $h(\k)$
as a differential operator except $G_0\ko$:
\begin{equation}
  G_1\ko h(\k) = -G_0\ko \gamma V_k \Dkx [ G_0\ko h(\k) ],
\end{equation}
\begin{equation}
  G_2\ko h(\k) = G_0\ko \gamma V_k \Dkx \left[ G_0\ko \gamma V_k \Dkx 
                 [ G_0\ko h(\k) ] \right].
\end{equation}
In contrast, $G^\dagger \ko$ acts on functions at the {\it left} side.

On the other hand,
the expectation value of supercurrent(\ref{current}) is written as the
thermal averaged form as
\begin{equation}
  \langle \j(\r) \rangle =e^* \sum_{k} \sum_{k'} \sum_\omega \sum_{\omega'} 
    {\rm tr} [R(\k,\omega;\k',\omega')(\hat{K}(\k)-e^* \hat{A})]
    \e^{i (k'- k ) \cdot r }
\end{equation}
where 
\begin{equation}
 \hat{K}(\k) = \pmatrix{ m_d^{-1} \k & m_v^{-1} \tilde{\k} \cr
                           m_v^{-1} \tilde{\k} & m_s^{-1} \k \cr} ,
 \tilde{\k} = (-k_x,k_y),
\end{equation}
\begin{equation}
 \hat{A} = \pmatrix{ m_d^{-1} \A & m_v^{-1} \tilde{\A} \cr
                         m_v^{-1} \tilde{\A} & m_s^{-1} \A \cr},
\tilde{\A} = (-A_x,A_y),
\end{equation}
and 
\begin{equation}
 R(\k,\omega;\k',\omega') = \langle \psi\kop \psi^\dagger\ko \rangle.
\end{equation}
Corresponding to the resolvent expansion(\ref{resolve}), 
the correlation function $R(\k,\omega;\k',\omega')$ is expanded as follows:
\begin{eqnarray}
  R(\k,\omega;\k',\omega') &=& G\kop \langle f\kop 
       f^\dagger\ko \rangle G^\dagger \ko \NY
 &=&  R_0(\k,\omega;\k',\omega') 
    + R_1(\k,\omega;\k',\omega') + R_2(\k,\omega;\k',\omega') + \cdots
\end{eqnarray}
where 
\begin{equation}
  R_0(\k,\omega;\k',\omega')
  = (2\pi)^4  G_0\kop
     a \delta(\k-\k') \delta(\omega-\omega') G_0^\dagger \ko,
\end{equation}
\begin{eqnarray}
  R_1(\k,\omega;\k',\omega') &=& (2\pi)^4
   [ G_1\kop a \delta(\k-\k') \delta(\omega-\omega') G_0^\dagger \ko \NY
  & & + G_0\kop a \delta(\k-\k') \delta(\omega-\omega') G_1^\dagger \ko],
\end{eqnarray}
\begin{eqnarray}
  R_2(\k,\omega;\k',\omega') &=& (2\pi)^4
   [ G_2\kop a \delta(\k-\k') \delta(\omega-\omega') G_0^\dagger \ko \NY
  & & + G_1\kop a \delta(\k-\k') \delta(\omega-\omega') G_1^\dagger \ko \NY
  & & + G_0\kop a \delta(\k-\k') \delta(\omega-\omega') G_2^\dagger \ko].
\end{eqnarray}

Now we can calculate supercurrent from the correlation functions.
The expressions for the supercurrent in each order are as follows:
\begin{equation}
\langle \j_0(\r) \rangle = e^* \sum_{k} \sum_{k'}
  \sum_\omega \sum_{\omega'} {\rm tr} [
  R_0(\k,\omega;\k',\omega') \hat{K}(\k) ]
  \e^{i (k'- k ) \cdot r }  ,
\end{equation}
\begin{equation}
\langle \j_1(\r) \rangle = e^* \sum_{k} \sum_{k'}
  \sum_\omega \sum_{\omega'} {\rm tr} [
   R_1(\k,\omega;\k',\omega') \hat{K}(\k) 
 - R_0(\k,\omega;\k',\omega') e^*\hat{A} ]
 \e^{i (k'- k ) \cdot r } , \label{j1}
\end{equation}
\begin{equation}
\langle \j_2(\r) \rangle = e^* \sum_{k} \sum_{k'}
  \sum_\omega \sum_{\omega'} {\rm tr} [
   R_2(\k,\omega;\k',\omega') \hat{K}(\k) 
 - R_1(\k,\omega;\k',\omega') e^*\hat{A} ]
 \e^{i (k'- k ) \cdot r }. \label{j2}
\end{equation}
Here we consider a thin film with thickness $d$, so that we calculate
the summation as integral with respect to $k_x$ and $k_y$,
and extract only the contributions from $k_z=0$.
Thus we have the expression for the paraconductivity from (\ref{j1})
\begin{equation}
  \sigma_{xx}=\frac{e^{*2} T}{d} \sum_{\omega} \sum_{k} {\rm tr}
 [ \{ -k_x^2 (G_0^\dagger W_x G_0 W_x G_0 \gamma G_0 + {\rm h.c.}) 
  + G_0^\dagger \gamma G_0^\dagger W_x G_0 \} (\gamma + \gamma^\dagger)] 
  \label{sxx}
\end{equation}
and for the excess Hall conductivity from (\ref{j2})
\begin{eqnarray}
\sigma_{yx} &=& 
  \frac{i e^{*3} T B}{d} \sum_{\omega} \sum_{k} {\rm tr} 
 [[ \{ k_y^2 W_y (1-2k_x^2 G_0 W_x) G_0 W_x G_0 (W_y G_0 \gamma + \gamma
    G_0 W_y ) -  {\rm h.c.} \} \NY
 & & -\{k_x^2 k_y^2 W_y G_0 W_x G_0 (W_y G_0 W_x G_0 \gamma
      + \gamma G_0 W_x G_0 W_y)  -  {\rm h.c.} \} \NY
 & & +\{k_x^2 k_y^2( W_y G_0^\dagger W_x G_0^\dagger W_y G_0 W_x G_0 \gamma
  -  {\rm h.c.} \} ] G_0 (\gamma + \gamma^\dagger) G_0^\dagger ]
  \label{syx}
\end{eqnarray}
where $G_0$ means $G_0 \ko$.

From the above expression 
we find that both $\sigma_{xx}$ and $\sigma_{yx}$
are even functions of $m_v^{-1}$ and
therefore we obtain $\sigma_{yy}=\sigma_{xx}$, $\sigma_{xy}=-\sigma_{yx}$; 
no anisotropy arises in the fluctuation
conductivity tensor in contrast to recent theoretical predictions
about vortex state below the critical temperature~\cite{Alv,Chang}.

Since the above expression for conductivity tensor is complicated,
it may be helpful to compare it with the former studies in the case of
$m_v^{-1}=0$. In this case the coupling of two components disappears
and the problem must reduce to that in the conventional single 
component superconductor,
which has been studied Aslamasov and Larkin(AL)~\cite{AL} and
Fukuyama, Ebisawa and Tsuzuki(FET)~\cite{FET}.
Then all matrices become diagonal and actually one can confirm
\begin{eqnarray}
  \sigma_{xx} &=& \frac{e^{*2} T}{d} \sum_{i=d,s} \sum_{\omega} \sum_{k}
    (\gamma_i+\gamma_i^*) [
   { -k_x^2 ( m_i^{-2} \gamma_i g_i^* g_i^3 + {\rm c.c.} )
      + m_i^{-1} \gamma_i g_i^{*2} g_i } ] \NY
  &=& \sum_{i=d,s} 
    \frac{e^2 T |\gamma_i|^2}{2\pi d \alpha_i \gamma_{i1}}
\end{eqnarray}
and
\begin{eqnarray}
  \sigma_{yx} &=& \frac{i e^{*3} T B}{d} \sum_{i=d,s} \sum_{\omega} \sum_{k}
    (\gamma_i+\gamma_i^*) [
    2       k_y^2 m_i^{-3} \gamma_i g_i^4 g_i^*  \NY
  & &  - 6 k_x^2 k_y^2 m_i^{-4} \gamma_i g_i^5 g_i^* 
   +   k_x^2 k_y^2 m_i^{-4} \gamma_i g_i^3 g_i^{*3} - {\rm c.c.} ] \NY
  &=& - \sum_{i=d,s}
    \frac{e^3 TB |\gamma_i|^2 \gamma_{i2}}
    {6 \pi d m_i \alpha_i^2 \gamma_{i1}^2},
\end{eqnarray}
where we introduced single component Green's functions
\begin{equation}
  g_i = \frac{1}{\i \gamma_i \omega + \eta_i}, 
  \eta_i = \frac{k^2}{2 m_i} + \alpha_i.
\end{equation}
and
\begin{equation}
  \gamma_{i1} = \re \gamma_i, \gamma_{i2} = \im \gamma_i
\end{equation}
These results correspond to AL and FET, respectively.

\begin{figure}
%\figureheight{10cm}
\begin{center}
\epsfile{file=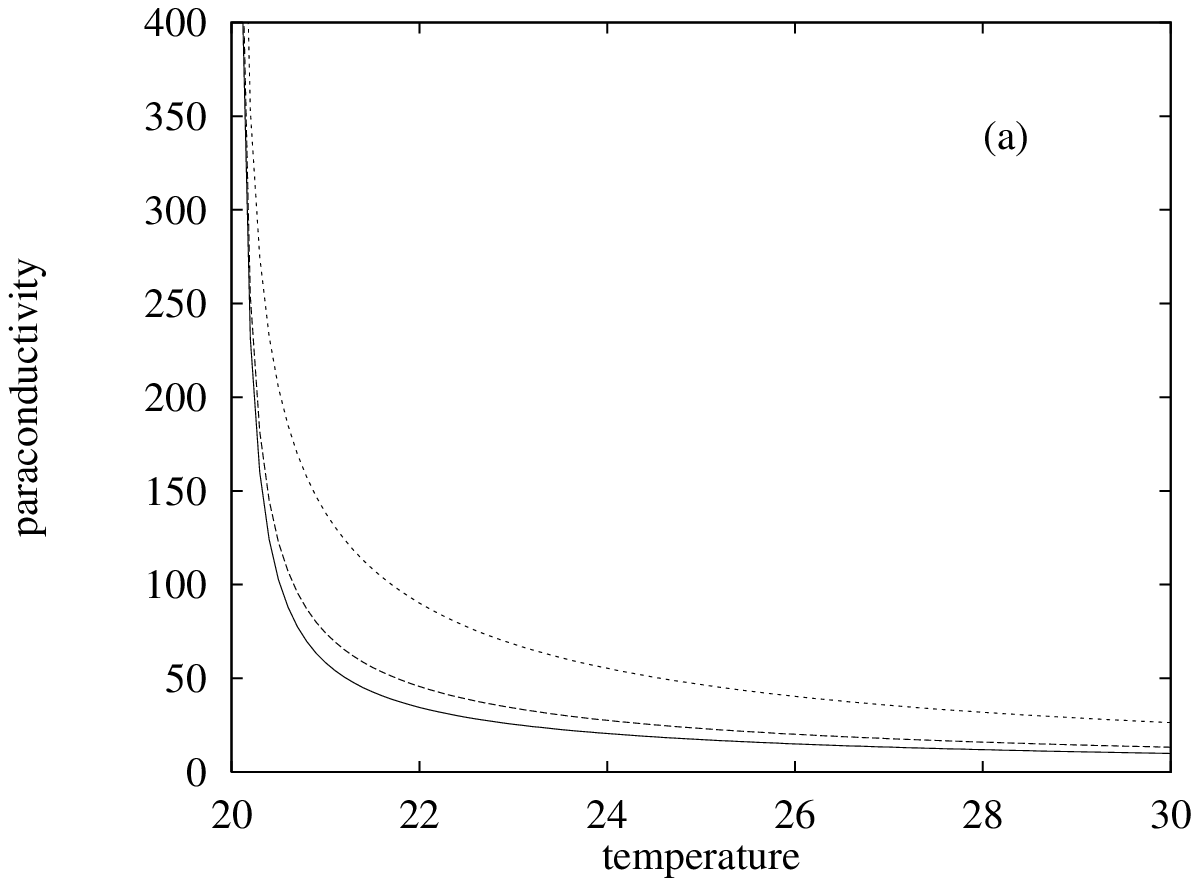,scale=0.6}
\end{center}
\begin{center}
\epsfile{file=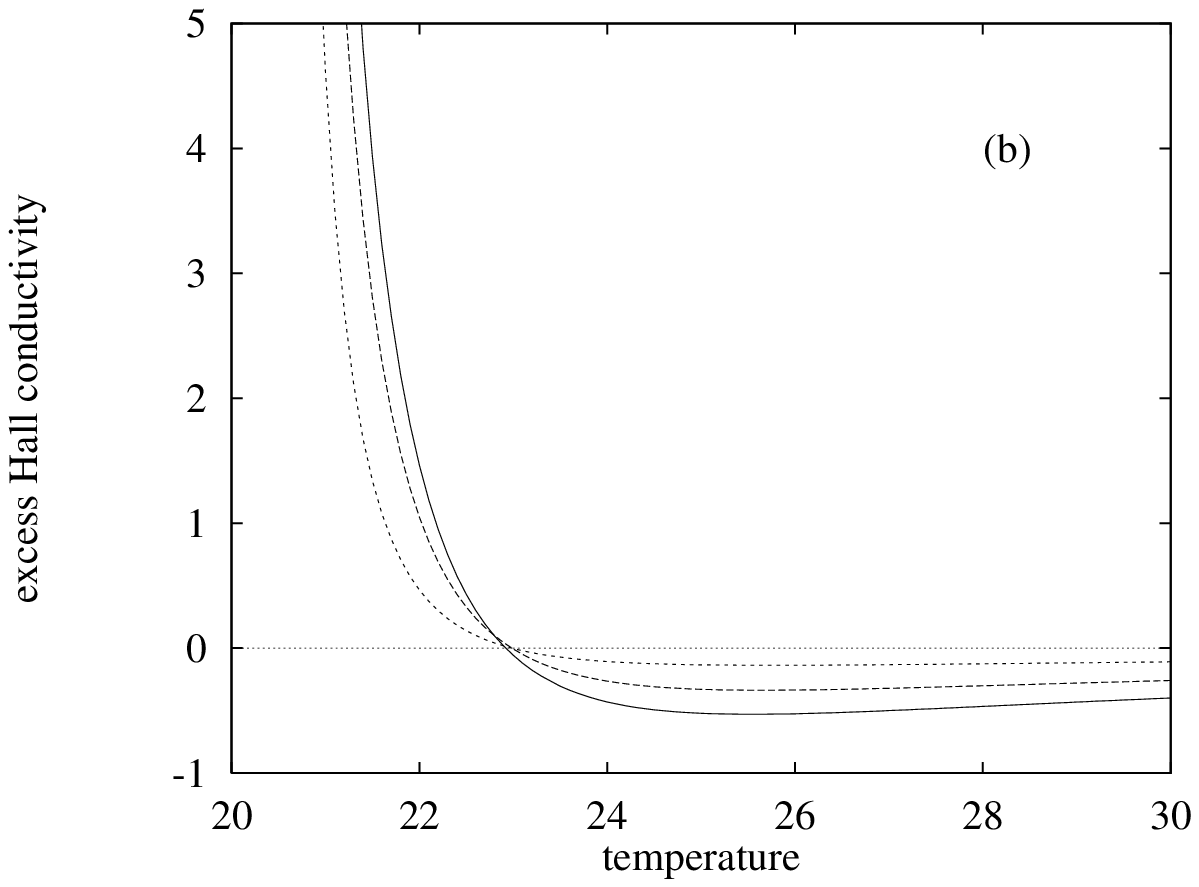,scale=0.6}
\end{center}
\caption{Temperature dependence of (a)paraconductivity $\sigma_{xx}$ and
(b)excess Hall conductivity $\sigma_{yx}$ for the coupling parameter
$m_v^{-1}=0$(solid line),$m_v^{-1}=3$(long dashed line),$m_v^{-1}=4$
(short dashed line) where
$T_d=20, T_s=16, m_d^{-1}=3,m_s^{-1}=6,\gamma_d=12+4\i,\gamma_s=4-10\i.$
Here we incorporated the temperature dependence by setting the GL
parameters $\alpha_d = T-T_d$ and $\alpha_s = T-T_s$.
}
\label{fig:1}
\end{figure}

\section{Numerical results} \label{sec3}
\begin{figure}
%\figureheight{10cm}
\begin{center}
\epsfile{file=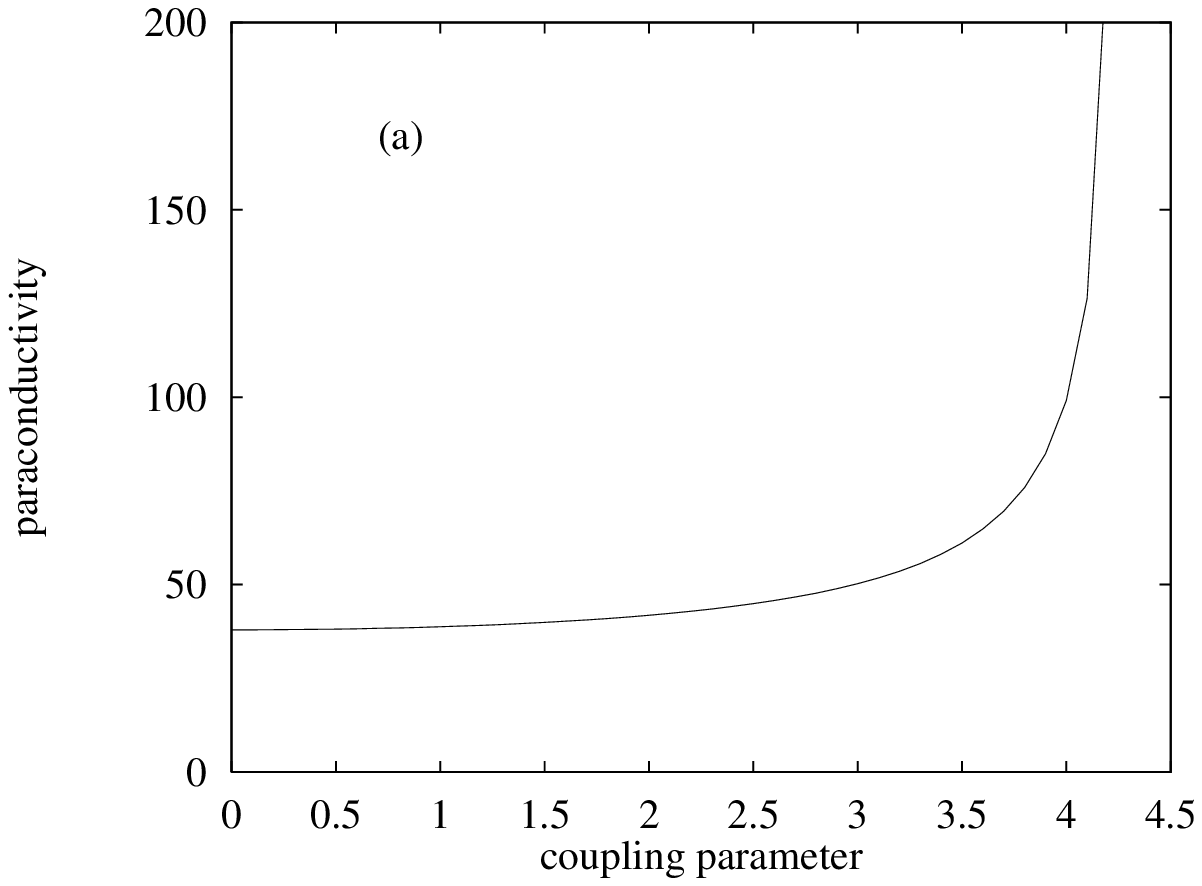,scale=0.6}
\end{center}
\begin{center}
\epsfile{file=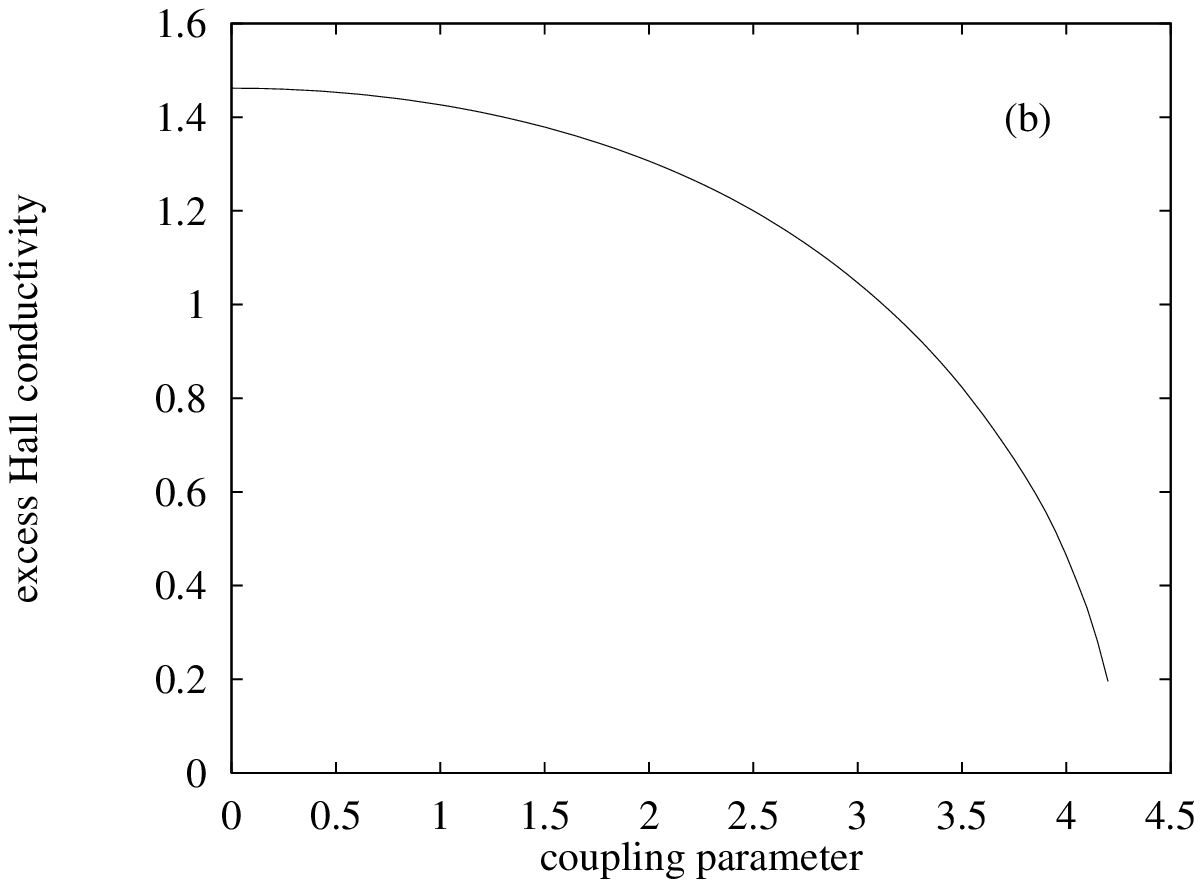,scale=0.6}
\end{center}
\caption{Coupling parameter dependence of (a)paraconductivity 
$\sigma_{xx}$ and (b)excess Hall conductivity $\sigma_{yx}$ at 
fixed temperature $T=22$ with the same set of parameters as in 
Fig.\ref{fig:1}.}
\label{fig:2}
\end{figure}

In this section we show the numerical results of paraconductivity(\ref{sxx})
and excess Hall conductivity(\ref{syx}). In Fig.1 we show the plot of
them as functions of temperature 
for various coupling parameter $m_v^{-1}$. 
Here the critical temperature of the $d$-wave component is chosen
to be higher than that of $s$-wave, and as a result, 
both $\sigma_{xx}$ and $\sigma_{yx}$ diverges at $T_d$ even in the
presence of the coupling of the two component order parameters.
It seems that $\sigma_{xx}$ is enhanced but $\sigma_{yx}$ is surpressed
due to the coupling of the two component order parameters.
In order to see the coupling effects more clearly,
we show the above results as functions of $m_v^{-1}$ in Fig.\ref{fig:2}.
In the light of GL free energy(\ref{GLF2}), $\det W_x > 0$ is required 
for the stability of the system against the spatial variation of 
order parameters.
This condition is reflected in the singularities at
$m_v^{-1}=\pm \sqrt{m_d^{-1}m_s^{-1}}$ in the fluctuation 
conductivity tensor.

\section{Summary and Discussions} \label{summary}
In this work we gave the formulation for the paraconductivity and
the excess Hall conductivity in the presence of the coupling of
two-component order parameter in the limit of weak external fields.
The phenomenological formulations based on the stochastic TDGL equation 
have been done by several authors~\cite{Schmid,Dorsey,dami}
and these correspond to AL process microscopically. 
Our calculation is considered to be their extension.
We also numerically calculated the dependences on
temperature and coupling parameter with the typical parameters
and found that the coupling of two components enhances 
paraconductivity but reduces excess Hall conductivity. 
Such tendency does not change even for the other set of parameters.

We paid attention to the stability condition about GL coefficients 
and saw the singular behavior of conductivity tensor
at critical value $m_v^{-1}=\pm \sqrt{m_d^{-1}m_s^{-1}}$, though
it could have not been attained if one had resorted to perturbative
expansions with respect to $m_v^{-1}$.

It is well known that excess Hall conductivity vanishes identically 
in the absence of
imaginary part of order parameter relaxation time within the conventional
framework of TDGL theory~\cite{FET}. On the other hand, 
several recent studies investigating vortex states based on two-component 
TDGL equations~\cite{Alv,Chang} have pointed out that the Hall effect is
also caused by the anisotropy of crystal lattice even if the relaxation
times are real. 
However, our calculation in the fluctuation regime resulted in
vanishing Hall conductivity in the case of $\im \gamma_d = \im \gamma_s = 0$,
and moreover, we found that no anisotropy arose in the conductivity 
tensor even for finite value of $m_v^{-1}$.
%This is because there is no preselected direction in the fluctuation
%of order parameter in contrast to the vortex lattice state.

Finally we mention an aspect of multiple transition in unconventional
superconductors from the viewpoint of superconducting fluctuation.
If the transition temperatures are close to one another in the presence of
the coupling between components,
the induction effect in superconducting fluctuation would gain its
significance. In this context,
the recent experiments through microwave surface impedance~\cite{sri}
which suggest multi-component superconductivity
can present an interesting system if those components 
correspond to order parameters of different symmetry.
To compare with such experiments, we need to carry out the extention
so as to include the frequency dependence in paraconductivity, 
which will be given elsewhere.

\section*{Acknowledgements}
The authours would like to thank M. Hayashi for fruitful discussions.
They are also indebted to discussions with K. Kuboki and H. Matsukawa.

\appendix
%\section{Calculation of the mean square values of two-component
%order parameters}


\begin{thebibliography}{99}
\bibitem{sri}
  H. Srikanth et al.: Phys. Rev. B{\bf 55} (1997) R14733;
  Phys. Rev. B{\bf 57} (1998) 7986.
\bibitem{will}
  M. Willemin et al.: Phys. Rev. B{\bf 57} (1998) 6137.
\bibitem{Joynt}
  R. Joynt: Phys. Rev. B{\bf 41} (1990) 4271.
\bibitem{Sigrist}
  M. Sigrist and T. M. Rice: Z. Phys. B{\bf 68} (1987) 9.
\bibitem{Xu}
  J. H. Xu, Y. Ren and C. S. Ting: Phys. Rev. B{\bf 52} (1995) 7663.
\bibitem{Xu2}
  W. Xu, Y. Ren and C. S. Ting: Phys. Rev. B{\bf 53} (1996) 12481.
\bibitem{Feder}
  D. L. Feder and C. Kallin: Phys Rev. B{\bf 55} (1997) 559.
\bibitem{Han}
  Q. Han and L. Zhang: Phys. Rev. B{\bf 56} (1997) 11942.
\bibitem{Franz}
  M. Franz, C. Kallin, P. I. Soininen, A. J. Berlinsky and A. L. Fetter:
  Phys. Rev. B{\bf 53} (1996) 5795.
\bibitem{Alv}
  J. J. Vicente Alvarez, D. Dominguez and C. A. Balseiro:
  Phys. Rev. Lett. {\bf 79} (1997) 1373.
\bibitem{Chang}
  D. Chang, C.-Y. Mou, B. Rosenstein and C. L. Wu:
  Phys. Rev. B{\bf 57} (1998) 7955.
\bibitem{Zap}
  M. Zapotocky, D. L. Maslov and P. M. Goldbart: Phys Rev. B{\bf 55}
  (1997) 6599.
\bibitem{kuboki}
  K. Kuboki and M. Sigrist: J. Phys. Soc. Jpn.{\bf 65} (1996) 361.
\bibitem{ichi}
  M. Ichioka, N. Enomoto, N. Hayashi and K. Machida:
  Phys. Rev. B{\bf 53} (1996) 2233.
\bibitem{hime}
  A. Himeda, M. Ogata, Y. Tanaka and S. Kashiwaya:
  J. Phys. Soc. Jpn.{\bf 66} (1996) 3367.
\bibitem{AL}
  L. G. Aslamasov and A. I Larkin: Phys. Letters {\bf 26A} (1968) 238.
\bibitem{FET}
  H. Fukuyama, H. Ebisawa and T. Tsuzuki: Prog. Theor. Phys. {\bf 46}
  (1971) 1028; H. Ebisawa and H. Fukuyama: Prog. Theor. Phys. {\bf 46}
  (1971) 1042.
\bibitem{naga}
  T. Nagaoka et al.: Phys. Rev. Lett. {\bf 80} (1998) 3594.
\bibitem{Schmid}
  A. Schmid: Phys. Rev. {\bf 180} (1969) 527.
\bibitem{Dorsey}
  S. Ullah and A. T. Dorsey: Phys. Rev. B{\bf 44} (1991) 262.
\bibitem{dami}
  D. C. Damianov and T. M. Mishonov: 
  Superlattices Microstruct. {\bf 21} (1997) 467.
\bibitem{spar}
  A. Kumagai and H. Ebisawa: unpublished.
\end{thebibliography}
\end{document}